\documentclass[pra,amssymb,amsmath,showpacs,superscriptaddress]{revtex4}
\usepackage{graphicx}

\begin{document}

\title{G{\"o}del Incompleteness and the Black Hole Information Paradox}
\author{R.     Srikanth}
\email{srik@rri.res.in}    \affiliation{Poornaprajna    Institute   of
Scientific Research, Devanahalli, Bangalore, India.}
\affiliation{Raman    Research   Institute,
Sadashiva Nagar, Bangalore,    India} 
\author{Srikanth Hebri}
\affiliation{Montalvo Systems, St. Marks Road, Bangalore, India.}

\begin{abstract}
Semiclassical reasoning  suggests that the process by  which an object
collapses into  a black hole  and then evaporates by  emitting Hawking
radiation may destroy information, a  problem often referred to as the
black hole information paradox.  Further,  there seems to be no unique
prediction  of where  the  information about  the  collapsing body  is
localized.  We propose that the latter  aspect of the paradox may be a
manifestation of  an inconsistent self-reference  in the semiclassical
theory of black  hole evolution.  This suggests the  inadequacy of the
semiclassical approach  or, at worst, that  standard quantum mechanics
and general  relavity are fundamentally incompatible.   One option for
the resolution for the paradox  in the localization is to identify the
G\"odel-like  incompleteness  that  corresponds  to an  imposition  of
consistency,  and introduce  possibly new  physics that  supplies this
incompleteness. Another option  is to modify the theory  in such a way
as to prohibit self-reference.   We discuss various possible scenarios
to  implement these options,  including eternally  collapsing objects,
black hole remnants,  black hole final states, and  simple variants of
semiclassical quantum gravity.
\end{abstract}

\maketitle

\section{Introduction \label{sec:srintro}} 
Attempts thus far to combine general relativity and quantum mechanics,
the  two  cornerstones of  our  description  of  nature, have  led  to
difficulties,  an  example of  which  is  the  black hole  information
parardox (BHIP) \cite{haw74,pre93}. BHIP  suggests that the process by
which an  object collapses  into a black  hole and then  evaporates by
emitting  Hawking radiation  is  not unitary.   The effect  apparently
leads  to  a  new  kind  of unpredictability,  quite  apart  from  the
conventional   one  associated   with  Heisenberg   uncertainty.   The
derivation of the paradox employs a semiclassical treatment of quantum
fields localized  close to  the event horizon  of a black  hole, which
would seem to leave open  the possibility of resolution through a more
detailed treatment of quantum gravity.  However, as the problem can be
posed of a region near the horizon  of a large black hole, it need not
invoke a  strong gravitational field, which suggests  that the problem
is amenable to a  local quantum field theoretic treatment \cite{boku}.
On  the other  hand, from  the string  theory standpoint  it  has been
argued that the detailed  knowledge of the Planck-scale physics cannot
be  ignored   even  if   there  is  no   strong  curvature   or  other
coordinate-invariant manifestation of the event horizon. Arguably, the
issue is  still open,  and continues to  attract efforts  at resolving
\cite{pre93,boku,hrv05,smo06,hormal03,gott03}    and   clarifying   it
\cite{bp07}.   In particular,  in the  loop quantum  gravity approach,
quantum effects eliminate black  hole singularities.  As a result, one
can  in  principle track  information  to  the  future of  a  would-be
singularity  \cite{smo2006}, thereby  preserving  information. It  has
also  been argued  that BHIP  may  be avoided  by attributing  Hawking
radiation solely to quantum  decoherence, considering that pure states
remain     pure     under     unitary,     closed-system     evolution
\cite{kie01,gam06}. This  is consistent  with the viewpoint  that pure
quantum states do not form black holes \cite{mye95}.

In the present work, we  propose that BHIP, in particular the question
of localization  of information in  an evaporating black hole,  may be
indicative   of  an  inconsistent   self-reference  occuring   in  the
semiclassical  treatment   of  black  hole   evolution  \cite{bhilos}.
Admittedly,  a   rigorous  study  of  this  claim   would  require  an
axiomatization of the  semiclassical theory.  Nevertheless, we believe
there  are plausible grounds  for believing  that there  are features,
presented here, that any such axiomatic theory should satisfy. Inspite
of the very abstract nature  of this approach to black hole evolution,
we will  be led below  to concrete, nontrivial consequences  for black
hole formation.   This work  may be primarily  regarded as a  plea for
injecting metamathematical considerations  in the study of fundamental
physics such  as quantum gravity,  and BHIP in particular.   A similar
case  can  be made  for  applying  quantum  information theoretic  and
computation theoretic insights to understanding the basic mathematical
structure of physical laws \cite{srigruska}.

The   remaining  article   is  arranged   as  follows.    In  Sections
\ref{sec:bhip} and  \ref{sec:godel}, we briefly review  the black hole
information  paradox  and   G\"odel's  first  incompleteness  theorem,
respectively.   The  ambiguity  in  the  localization  of  information
falling  into  an evaporating  black  hole  is  introduced in  Section
\ref{sec:bhiloc}.  An argument that  the localization problem may be a
formal  inconsistency  in the  semiclassical  theory  is presented  in
Section \ref{sec:sriambi},  and that this inconsistency  can be viewed
as    self-referential   in   origin    is   presented    in   Section
\ref{sec:srincons}.  The question of restoring consistency by invoking
various  ways to account  for the  G\"odel incompleteness  obtained by
imposing  consistency  is  considered in  Section  \ref{sec:srincomp}.
That of  restoring consistency by means of  avoiding self-reference is
considered in Section \ref{sec:avsf}.  Finally, we conclude in Section
\ref{sec:konklu}.

\subsection{The black hole information paradox \label{sec:bhip}}
A brief  introduction to BHIP is  as follows.  We denote  by $H_M$ the
Hilbert  space of  a  collapsing  body $M$,  of  dimension $N$,  where
$N=e^{\cal S}$,  and ${\cal S}$ is  the black hole's  entropy.  In the
semiclassical   treatment  of  quantum   field  fluctuations   on  the
background spacetime  determined by the collapse and  evaporation of a
black hole,  the Hilbert  space of the  fluctuations can  be separated
into two  subsystems, given  by Hilbert spaces,  respectively, $H_{\rm
in}$ and  $H_{\rm out}$ (each  also of dimension $N$),  located inside
and outside the  horizon.  The correlations between the  two fields is
characterized  by  the Unruh  quantum  state  $|\Phi\rangle_{\rm in  +
out}$,  which looks  like  the vacuum  in  the far  past, a  maximally
entangled pure state \cite{gott03}
\begin{equation}
|\Phi\rangle_{\rm in + out} = \frac{1}{\sqrt{N}}\sum_{j=1}^N
|j\rangle_{\rm in}|j\rangle_{\rm out},
\label{eq:unru}
\end{equation}
where $|j\rangle_{\rm  in}$ and $|j\rangle_{\rm  out}$ are orthonormal
bases  for $H_{\rm  in}$ and  $H_{\rm out}$,  respectively.  The Unruh
state contains a  flux of particles in $H_{\rm  out}$, that constitutes
the Hawking radiation.

To the  outside observer $H_{\rm  in}$ is inaccessible, and  the field
localized   outside  is   in   the  maximally   mixed  state   $\sum_j
(1/N)|j\rangle_{\rm out}\langle  j|_{\rm out}$ containing  no detailed
information  about  $M$.   When   back-reaction  is  included  in  the
semiclassical approximation, the black  hole will slowly lose its mass
through Hawking  radiation, and  disappear.  From the  classical black
hole geometry, the information about what formed the black hole cannot
come out  without violating causality. So  at late times,  we obtain a
mixed state,  even though  $M$ began in  a pure state.   Clearly, this
process cannot be described as  a unitary evolution, and suggests that
black holes destroy information.  This problem is often referred to as
BHIP. However, it  is convenient for us to regard it  as one aspect of
the  full  paradox,  which  (aspect)  we shall  call  the  black  hole
information  loss problem.  There  is another  aspect of  the paradox,
introduced in  Section \ref{sec:bhiloc}, which we call  the black hole
information localization problem.

\subsection{G\"odel Incompleteness\label{sec:godel}}
A  formalization  or axiomatization  of  arithmetic  (in general,  any
deductive theory)  is the  reduction of arithmetic  to a small  set of
initial formulas and rules of symbolic manipulation, such that a chain
of formulas obtained by  manipulation in the formal system corresponds
to  and  represents  deductions  in  arithmetic.  By  looking  at  the
correspondence between  the formal system and the  deductive theory in
reverse, Hilbert originated metamathematics, his name for the study of
rigorous  proof in  mathematics  and symbolic  logic.   Here a  formal
system is  a `game' constructed, independently  of its interpretation,
as a sequence of formulas obtained mechanically according to the rules
of  symbolic manipulation,  starting from  the initial  formulas.  The
formal system  is interpreted as representing the  deductive system if
the initial  formulas can be  interpreted as expressing the  axioms of
the theory, and the rules  of symbolic manipulation, its logical rules
of inference.  Then, a metamathematical proof that a formula occurs in
a sequence  of formulas of the  formal system yields a  proof that the
proposition which is  the interpretation of this formula  is a theorem
of the deductive theory.

From  the   standpoint  of  mathematic  logic,  it   is  important  to
distinguish   between   statements   in   a  deductive   theory   from
meta-statements in the metatheory,  which studies concepts, proofs and
truths  in the theory.   Failure to  do so  can lead  to inconsistency
through self-reference,  of which a  well-known example is  the liar's
paradox: ``this statement  is false".  Here the statement  acts as its
own metastatement.   If the statement is  true, then it  is false, and
conversely:  a contradiction.   From a  syntactic viewpoint,  a formal
system is consistent  if for any proposition $\alpha$,  at most one of
$\alpha$ and its negation $\neg\alpha$ is provable.  The formal system
is complete if for any  proposition $\alpha$, at least one of $\alpha$
and $\neg\alpha$ is provable \cite{semantic}.

G\"odel's (first) incompleteness  theorem, perhaps the most celebrated
result in metamathematics,  states that any formal system  that is (1)
rich enough  to encompass arithmetic,  (2) is finitely  specified, and
(3) consistent, contains a proposition  that can neither be proved nor
refuted within  the system \cite{God},  and is thus  incomplete.  Here
`finitely  specified' means  that there  is an  algorithm to  list all
axioms (initial  formulas) and rules of inference  (rules for symbolic
manipulation),  which  may  be  countably  infinite.   Regarding  (3),
G\"odel    actually    requires     the    stronger    condition    of
$\omega$-consistency  \cite{Omega},  a subtlety  we  may ignore  here.
Every  deductive  theory  that  includes  elementary  arithmetic  (the
notions  of natural  numbers, and  of the  operations of  addition and
multiplication) also inherits this incompleteness.  Only theories with
sufficiently  simple  logical  structure,  such  as  propositional  or
sentential  calculus, Presburger  arithmetic and  elementary geometry,
are complete.

G\"odel's theorem  is a  consequence of the  fact that  arithmetic has
enough  expressive power  to  allow meta-arithmetic  statements to  be
mirrored into it, thus making some sort of self-reference unavoidable.
Crucial to  G\"odel's proof is the  observation that the  symbols of a
formal arithmetic system, and hence formulas and proofs constructed in
it, can  be assigned a unique  number, now called  the G\"odel number.
Any other method  of assigning numbers to these  objects in one-to-one
fashion  will also  work.  As  a result,  meta-arithmetical statements
about arithmetic can be paraphrased arithmetically as statements about
their G\"odel numbers.  The  meta-arithmetic statement that a sequence
$\alpha$  of formulas  is  a proof  of  the formula  $\beta$, or  that
formula $\gamma$  is provable, can  be expressed, respectively,  as an
arithmetical  relation between  the G\"odel  numbers for  $\alpha$ and
$\beta$, or an arithmetic property  of the G\"odel number of $\gamma$,
and thus expressed  in the formal system.  This  isomorphic mapping of
meta-arithmetic into arithmetic opens up the danger of self-reference.
If one takes care to set  up blocks that prohibit inconsistency of the
liar's paradox type, one is left with incompleteness as a side-effect,
as it were.

Let us briefly present a simplfied, illustrative but unrigorous sketch
of G\"odel's  proof.  Let  system $P$ be  a formalization  of ordinary
arithmetic,  whose alphabet ${\bf  P}$ consists  of the  symbols ``0''
(zero), ``$s$''  (successor), the logical  constants $\forall$, $\neg$
(negation), $\vee$, and variables of the first type $x_1, y_1, \cdots$
(for individuals,  the numbers including  0), variables of  the second
type  $x_2,y_2,\cdots$  (for  classes  of  individuals),  and  so  on.
Metamathematically,  it  is  immaterial  what  symbols  we  choose  to
represent these  basic signs,  and we may  choose natural  numbers for
them.  Accordingly,  a formula is  a finite series of  natural numbers,
and  a particular  proof is  a  finite series  of a  finite series  of
natural numbers.   Metamathematical concepts and  propositions thereby
become  concepts  and  propositions  concerning natural  numbers,  and
therefore,  at  least partially  expressible  in  the  symbols of  $P$
itself.   In  particular,  G\"odel  shows  that  the  metamathematical
concepts    ``formula'',   ``axiom'',''variable'',   ``proof-schema'',
``provable formula'', etc., are definable within $P$.

We  call formulas  involving  a single  variable  as class-signs.   If
$\alpha$ is a class-sign, and  $t$ a number, we designate by $[\alpha;
t]$ the formula obtained by substituting  the sign for $t$ in place of
the  free  variable in  $\alpha$.   Let  every  class-sign be  somehow
ordered, e.g., lexicographically.  The concept class-sign and ordering
$R$ can be defined in $P$.  Let $R(n)$ denote the $n$th class-sign.

We define a set $K$ of whole numbers by: 
\begin{equation}
\label{eq:clasgn}
n \in K ~\equiv~ [R(n);n] {\rm ~is~}{\rm not~}
{\rm provable~in~} P.
\end{equation}
As the r.h.s of Eq.  (\ref{eq:clasgn})  is definable in $P$, so is the
concept $K$ in  the l.h.s.  That is, there is a  class-sign $W$ in $P$
such that $[W;n] \equiv n \in K$.  For some positive integer $q$, $W =
R(q)$. We will find that the string, a G\"odel sentence for $P$,
\begin{equation}
\label{eq:God}
[R(q);q],
\end{equation}
is  {\em  undecidable} in  $P$.   If  proposition (\ref{eq:God})  were
provable  in  $P$,  then  so  would  the  proposition  $q  \in  K$  by
definition. The latter would imply $[R(q);q]$ is {\em not} provable in
$P$ according to Eq.  (\ref{eq:clasgn}).  This is a contradiction.  On
the other  hand, if proposition (\ref{eq:God}) were  refutable in $P$,
i.e., $\neg[R(q);q]$ were provable in $P$, this would supply the proof
that $\neg(q \in K)$, so that, by Eq. (\ref{eq:clasgn}), $[R(q);q]$ is
provable  in  $P$.   Again,  we obtain  a  contradiction.   Therefore,
assuming  $P$   is  consistent,  $[R(q);q]$  is   undecidable  in  $P$
\cite{sriform}.  Thus $P$  is incomplete.  Proposition (\ref{eq:God}),
which  involves supplying  a  formula  its own  serial  number as  its
argument,  is  an  instance  of  the  diagonal  argument  \cite{linz},
pioneered  by  the  mathematician Cantor  \cite{sricantor}.   Clearly,
(\ref{eq:God}) is true,  since if it were false,  it would be provable
in  $P$,  thereby contradicting  itself.   We  thus  have the  curious
situation that (\ref{eq:God}) is  known to be true metamathematically,
even though it is unprovable in $P$ \cite{add}.

An existential proof  of G\"odel's theorem is obtained  by noting that
the  set  $\Pi$  of   provable  propositions  in  a  formalization  of
arithmetic  is  recursively  enumerable  (r.e.)  \cite{re},  in  fact,
recursive  \cite{rec}, whereas the  set $T$  of truths  expressible in
arithmetic  is not  r.e  \cite{usp}.  In  a (semantically)  consistent
formalization,  clearly, $\Pi  \subseteq T$.   Since $T$  is  not r.e,
there should be truths that  are unprovable in the given formalization.

G\"odel's incompleteness theorem is related to Turing uncomputability,
the     unsolvability    of    certain     problems    algorithmically
\cite{tur36,sricantor}.  If every  proposition in an arithmetic system
$P$ were decidable by an algorithm  $G$, this could serve as the basis
to solve the halting problem for Turing machines, which is known to be
undecidable  \cite{halt}.  The  unsolvability of  the  halting problem
thus implies the existence  of undecidable propositions in $P$. 

Turing   machines  and   the  system   $P$  derive   their   power  of
self-reference  from their  universality: the  existence  of universal
Turing machines in  the case of the former, and  the ability to mirror
meta-arithmetic statements in the case of the latter.  However we will
find that  one can construct  simpler systems in  which self-reference
occurs, leading  to incompleteness  or inconsistency.  As  an informal
example,  we  note disentangling  proposition  (\ref{eq:God}) that  it
asserts its own unprovability in $P$.   It may thus be regarded as the
consistent,  and  hence incomplete,  version  of  the liar's  paradox,
which, by asserting its own falsehood, is complete, but inconsistent.

\section{Black hole information localization problem \label{sec:bhiloc}}

A  problem in  BHIP closely  related to  the information  loss problem
concerns the situation that as a radiating semiclassical Schwarzschild
black hole shrinks,  and possibly fully evaporates, there  seems to be
no unique prediction of where  the information about $M$ is localized.
The event horizon being a globally defined property of a spacetime, to
a  freely falling observer  (called Alice),  matter falling  towards a
black hole encounters nothing unusual while crossing the horizon.  She
finds the quantum information contained in the initial matter $M$ pass
freely into the interior of the black hole.  In contrast, according to
an observer Bob, who is outside the event horizon, at a large distance
from the black hole and approximately  at rest with respect to it, the
collapsing or  infalling object appears to increasingly  slow down and
to freeze asymptotically at the  horizon. He finds that it never quite
crosses the event horizon  during the semiclassical stage, and perhaps
even  later, as the  black hole  evaporates, possibly  entirely.  This
lack   of   a  unique   prediction   of   the   localization  of   the
infalling/collapsing matter  in the  semiclasscal theory is  the black
hole  information localization problem.  This can  be made  clearer as
follows.

\subsection{The black hole information localization problem viewed 
as formal inconsistency \label{sec:sriambi}}

We consider Alice  falling towards a Schwarzschild black  hole of mass
$m > 0$.  To Bob, the  coordinate time $t$ of the Schwarzschild metric
corresponds approximately to his  proper time.  Alice initially stands
close  to Bob  before  propelling herself  forward  and then  allowing
herself to fall  freely into the black hole.  The  black hole mass $m$
is  assumed to  be sufficiently  large that,  from her  viewpoint, all
events in her worldline segment  up to her infall into the singularity
can be regarded to good accuracy as happening in a region of spacetime
endowed  with a  classical, time-independent  metric.   Let $\epsilon$
denote the event of this  worldline of Alice intersecting the horizon.
Consider      the       Kretschmann      scalar      $K^A(\tau)      =
R_{abcd}(\tau)R^{abcd}(\tau)$,  where $R_{abcd}(\tau)$ is  the Riemann
tensor  along  Alice's  worldline,  parametrized by  her  proper  time
$\tau$.   For  convenience, we  set  $\tau=0$  at  $\epsilon$.  For  a
Schwarzschild  metric  (\ref{eq:schs}), $K^A=48m^2/r^6$  \cite{hen00}.
In  particular,  event  $\epsilon$  is  marked  by  the  scalar  value
$K^A(0)=3/4m^4$.

According to Bob, because  of gravitational redshift, Alice is falling
ever more slowly, but never  quite getting to the horizon.  Throughout
the  semiclassical regime, we  may assume  that the  evaporating black
hole  is approximately static  and spherically  symmetric, with  a the
shrinking  horizon,  and  that  Bob remains  the  distant,  stationary
observer.  Thus, even when substantial black hole mass has evaporated,
Alice will  not yet have crossed  through the horizon,  as viewed from
his perspective.  A  quantum gravity scenario in which  she never does
so as seen by Bob is  not inconceivable.  Indeed, this is the accepted
situation in string theory  \cite{boku}.  Even if Alice's crossing the
horizon  does eventually  happen,  this event  (denoted $\eta$)  would
presumably have to  occur in the strongly quantum  gravity regime.  We
expect that  the Kretschmann  scalar $K^B(0)$ at  $\eta$ would  be far
larger.  Even if $K^A(0) = K^B(0)$ by an extraordinary conspiracy, the
functions $K^A(\tau)$ and $K^B(\tau)$ could  not be the same for $\tau
\ge 0$, as  the former occurs in a  purely classcal spacetime, whereas
the  latter in a  full quantum  gravity regime.   In a  scenario where
Alice  does not  cross the  horizon  in Bob's  perspective, $\eta$  is
termed a ``null event''.

Either way, we are led to  conclude that $\epsilon \ne \eta$, and that
Alice's  and Bob's  perspectives  are mutually  incompatible.  To  the
extent that the assumptions made  are realistic, this inability of the
semiclassical  theory  to assign  a  unique  spacetime  location to  a
physical event  \cite{nota} may be regarded as  a formal inconsistency
in the theory.   In particular, let ${\bf g}$  be the proposition that
the quantum information pertaining to an infalling body passes through
the  horizon  at  event  $\epsilon$.  From  Alice's  perspective,  one
predicts ${\bf g}$, and from Bob's perspective, one predicts $\neg{\bf
g}$.  The inconsistency is that the theory does not (seem to) assign a
unique destiny to the infalling information.

It is  worth stressing that  this incompatibility is  fundamental, and
should not  be thought  of as a  manifestation of  non-Boolean quantum
logic defined in the state  space $H_M \otimes H_A \otimes H_B$, where
$H_A$ and $H_B$ are the Hilbert spaces of Alice and Bob, respectively.
In  particular, the  incompatibility  cannot be  accounted  for by  an
entanglement between  the black hole, Alice and  Bob, where $\epsilon$
($\eta$) occurs relative to Alice (Bob). The simple reason for this is
that  distant  Bob  need  not  interact,  and  thus  need  not  become
entangled, with  $M$ and Alice.  But  the basic reason is  that such a
`relative state' interpretation is possible  only if Alice and Bob are
two  states  of the  {\em  same}  system,  rather than  two  different
systems, as is the case here \cite{2sys}.

In  retrospect, the  inconsistency stems  from the  fact  that general
relativity  permits the  co-existence of  observers of  very disparate
powers:  on  the one  hand,  Bob,  and on  the  other,  Alice, who  is
infinitely more powerful  than him in the sense  that there are events
(such  as $\epsilon$)  to his  infinite future  that happen  in finite
proper  time according  to Alice  (but none  vice-versa).  In  fact, a
disparity of  this kind may  serve as a  basis to construct  a general
relativitistic hypercomputer to  compute Turing uncomputable functions
\cite{maho}.  The  origin of the localization problem,  then, seems to
be the combined  effect of the existence and  diminution of the domain
of the  `infinitely powerful' observer,  given by the interior  of the
Schwarzschild black hole.

To our  knowledge, of the  various efforts to resolve  BHIP, currently
string  theory alone  seems to  acknowledge the  localization problem.
According to  this view, an  infalling object carries  its information
intact  through  the horizon  as  seen  in  Alice's perspective.   Bob
perceives  the black  hole draped  by a  heated membrane  that  is the
source of  the Hawking  radiation, and situated  just above  the event
horizon.  He  considers any infalling information  to become disrupted
upon  approaching this membrane,  and re-emitted  as radiation  to the
exterior  universe,  keeping the  late-time  state  pure.  While  this
offers the prospect  of solving the information loss  problem, that of
information localization still remains.

The standard understanding among  string theorists \cite{boku} is that
this perspective-dependent nonlocal  existence of infalling matter can
be  accounted  for by  the  principle  of  black hole  complementarity
\cite{suss},  the idea  that Alice's  description is  complementary to
Bob's,  somewhat in the  same way  that the  description of  a quantum
particle  in  terms  of   position  and  momentum  are  complementary.
Historically, this  notion of nonlocality  has served to  motivate the
holographic  principle  \cite{hooft}.   Thus  the  incompatibility  of
Alice's  and Bob's  perspectives  is  treated as  a  new principle  of
relativity, rather than as an inconsistency.  

The  idea that  the observation  or  non-observation of  an event  can
depend  on the  the  observer's  reference frame  occurs  also in  the
context the  Unruh effect \cite{unru76}, where the  Unruh radiation is
observed in an accelerated frame but not in the corresponding inertial
frame.  Indeed, the existence of  Unruh radiation can be linked to the
apparent  event horizon  perceived  by an  accelerated observer,  thus
putting  it in  the  same conceptual  framework  as Hawking  radiation
\cite{hara}. It has  long been argued that the  appearance of particle
events is  observer dependent \cite{ful73}.  Thus,  the observation or
non-observation of  an event of  horizon-crossing or not  crossing can
reasonably  depend  on  the  frame   of  the  observer  and  does  not
necessarily signal inconsistency.

In string theory, this remarkable position is justified by appeal to a
verificationist philosophy: one  can choose not to be  bothered by the
`cloning' of information because it cannot be verified to have occured
by  an  observer  within  the semiclassical  effective  theory  regime
\cite{suss}, inasmuch as any attempt to do so would require energy far
beyond the Planck scale.  Thus,  this standpoint defers a treatment of
the problem from the semiclassical  regime to a full theory of quantum
gravity.

\subsection{BHIP viewed as an inconsistent self-reference in quantum gravity
\label{sec:srincons}}
Since physics  is described  in the language  of mathematics, it  is a
deep  yet natural  question to  ask  what, if  any, is  the impact  of
G\"odel  incompleteness  and  Turing  uncomputability  on  physics,  a
question       that      has      elicited       varied      responses
\cite{cas96,svo03,perzur,haw02}.  Recently, Heisenberg uncertainty has
been related \cite{cal04} to G\"odel incompleteness through the notion
of  information theoretic  incompleteness \cite{cha}.   An interesting
survey of the possible  impact of G\"odel's incompleteness theorems on
physics may be found in Ref. \cite{bar02}.

We expect that  the laws of physics possess  sufficiently rich logical
structure to  express elementary arithmetic, as  evidenced for example
by  the simple  fact of  existence of  electronic computers.   Hence a
formalization  of  physics should  be  able  to express  metatheoretic
propositions,  such as  provability  in itself.   If consistent,  this
formal  system  (and  by  extension,  physics  itself)  will  then  be
incomplete.   Intuitively, we regard  physical systems  as `computing'
their own  evolution \cite{llo00}, and  thus believe that there  is an
algorithm  to compute  any property  of a  system.   The Church-Turing
thesis of computer science \cite{linz}  may then be invoked to suggest
that no physical  process, interpreted as a procedure  for computing a
function,  can  be  more  powerful  than Turing  machines.   Thus  the
incompleteness   of   physics   could   manifest   in   the   physical
uncomputability of Turing uncomputable problems \cite{sri06}.

On the other hand, a candidate theory of physics could be inconsistent
through  self-reference.   It may  not  be  straightforward to  detect
inconsistency   in  the   laws  of   the  theory.    G\"odel's  second
incompleteness theorem shows that the  formal system $P$ can prove its
own consistency if and only if it is inconsistent \cite{go2}.  One way
to deal with the issue of consistency of a physical theory would be to
axiomatize  it  and  then   try  to  demonstrate  its  consistency  or
inconsistency, failing which  one may hope that it  is consistent; or,
one may try  to prove its consistency metatheoretically.   Even in the
absence of  a formalization of  the theory, an inconsistency  might be
revealed  through  the  prediction   of  some  genuine  paradox.   The
detection of an inconsistency would  imply that the theory in question
is inadequate, or, what is less likely, that Nature herself harbors an
inconsistency  \cite{bar02}.  Here  we propose  that  the localization
paradox  of BHIP  may  signal an  inconsistent  self-reference in  the
semiclassical approach to black hole evolution.

To view the  formal inconsistency of the BHIP  localization problem as
self-referential, it may be re-cast  as a time paradox.  Suppose Alice
freefalls for a while, but  before reaching the horizon, switches on a
rocket  and returns  back to  Bob.  In  this case,  no  paradox arises
because  Alice's  and  Bob's  perspectives will  coincide,  and  their
respective observations  will be continuously  transformable into each
other, within the semiclassical  theory.  However, if Alice chooses to
continue to freefall,  and enters the black hole  at event $\epsilon$,
then Bob's  perspective does not register $\epsilon$,  but instead the
incompatible (possibly null) event $\eta$.

\begin{figure}
\includegraphics[width=7.0cm]{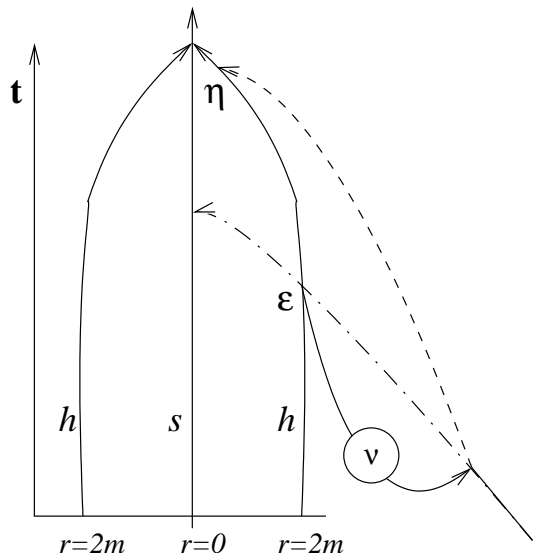}
\caption{Bird's eye `meta-view' of  the BHIP localization paradox as a
self-referential classical  information circuit. Negative  time travel
is  indicated through the  curve labeled  by the  encircled ``$\nu$''.
(a) The lines $s$ and  $h$ represent the singularity and the shrinking
horizon,  respectively.  The  dash-dotted  and dashed  curves are  the
worldlines  of an  infalling object,  according to  Alice's  and Bob's
perspectives,  respectively.   We note  that  $\eta$,  if it  happens,
presumably occurs in the fully quantum  gravity regime, and if it is a
null event, the `$\eta$-worldline' will not intercept the horizon. The
negative-time feed-back  is triggered if and only  if event $\epsilon$
happens.  Prior to the  feed-back induced nonlocal split, the object's
worldline exists in an unambiguous past.}
\label{fig:sriganesh0}
\end{figure}

This  situation may be  described in  the following  somewhat fanciful
language.  The information about Alice (or $M$) propagates towards the
black hole initially from an unambiguous past.  If Alice does not fire
her rocket  but freefalls into  the black hole, a  signal time-travels
from her future self at the event $\epsilon$ to her past self occuring
earlier, instructing the latter to shift to the incompatible worldline
leading to  $\eta$ in Bob's  perspective.  Thus, if an  infalling body
enters the  black hole at $\epsilon$,  it `will not  have entered' the
black hole  during that  event.  And  if $\eta$ is  a null  event, the
paradox is  that if the body enters  the black hole, then  it will not
have entered.   In this  sense, the BHIP  localization problem  may be
viewed as  a physical version of  the liar's paradox.   If Alice fires
her  rocket to  return  back  to Bob,  no  such time-traveling  signal
occurs.  More  generally, this signal  is generated, and  an infalling
object `experiences'  a nonlocal  splitting of the  self into  the two
mutually  incompatible perspectives,  if and  only if  it  crosses the
horizon  in  Alice's perspective.   Since  this time-traveling  signal
occurs  across  the  perspectives,  the self-reference  is  perceived,
strictly  speaking, in the  `meta-perspective' that  has a  bird's eye
view of both perspectives, rather than in the Alice perspective or Bob
perspective alone (cf.  Figure \ref{fig:sriganesh0}).

Although  this self-reference  is tied  in  a complicated  way to  the
causal  structure  of  spacetime  in general  relativity  modified  by
quantum mechanics,  the essential idea  of the inconsistency  as being
due to  a temporal self-reference, and of  the incompleteness obtained
by imposing  consistency, can be  roughly demonstrated using  simple
`self-referential  circuits' that compute  a one-bit  partial function
(cf.  Appendix \ref{sec:sritysk}).

A quick way to impose  consistency in the BHIP localization problem is
to proscribe objects from falling  into the inconsistent zone which is
the evaporating black hole.  Applied to any infalling body, this would
suggest  that the  horizon never  forms in  finite time  in  the first
place.  G\"odel incompleteness would  then correspond to the situation
that,  if the theory  is consistent,  it could  somehow not  allow the
dynamic formation of a solution (the Schwarzschild black hole) that it
nevertheless allows to exist, because  if this solution were formed in
finite time,  it would be  inconsistent.  One could then  require that
the detailed dynamics of  the infalling matter, possibly involving new
physics, would somehow conspire to prevent an object's collapsing into
the  horizon. We  will return  to  this point  in detail  in the  next
Section, where we consider this and various other proposals to resolve
BHIP in this light.

\section{Towards resolving BHIP via G\"odel incompleteness
\label{sec:srincomp}}  
Even  if  we  admit  that  the  semiclassical  theory  of  black  hole
evaporation may  be inconsistent  in the above  sense, nonrelativistic
quantum  mechanics  and  classical  general  relativity  are  arguably
consistent in their  own domains. 

This suggests that the axioms  of quantum mechanics are not compatible
with those of general relativity, and that BHIP may be a manifestation
of  this  incompatibility.   Again,  in  the  absence  of  a  rigorous
axiomatization  of  semiclassical   general  relativity,  three  broad
operational responses to  the situation may be considered  in order to
eliminate the inconsistency, either  by averting self-reference, or by
invoking  possibly new  physics  that would  explain the  G\"odel-like
incompleteness corresponding  to an imposition of  consistency: (A) to
somehow  thwart  the full  collapse  of $M$  into  a  black hole  from
happening in finite time; (B) to modify the semiclassical theory, with
the modifications being  understood as coming from the  full theory of
quantum  gravity;  (C)  to  modify standard  non-relativistic  quantum
mechanics, and/or classical general relativity.

The first two options are considered sequentially in the following two
subsections.  Option  (C) is considered  in the next  Section.  Option
(A) is  concerned with the  introduction of G\"odel  incompleteness in
the form of  prohibiting the formation of black  holes in finite time.
Option (B)  admits the  inconsistency in the  initial phase  of infall
through the horizon, but enforces late-time consistency, by means of a
black hole  remnant or a black  hole final state.  Option  (C) aims to
modify one or both of the ingredient theories of semiclassical quantum
gravity, i.e.,  quantum mechanics and  general relativity, so  that no
BHIP-like self-reference occurs in putting them together.

\subsection{Eternally collapsing objects instead of evaporating 
black holes.\label{sec:ever}}

As  noted  briefly  at  the  end of  Section  \ref{sec:srincons},  the
simplest  way of imposing  consistency on  the evolution  of infalling
bodies implies  that non-zero  mass black holes  should never  form in
finite time.   If we believe  in the consistency of  the semiclassical
approach, we may then `predict' the existence of a dynamical mechanism
that would explain how a body  may be prevented from collapsing into a
black hole  of non-zero mass  in finite time.   The physics of  such a
mechanism  would supply  the  required G\"odel  incompleteness. It  is
possible that we would require new physics to fulfil this purpose, but
we  expect that  there would  be little  departure  from semiclassical
theory near the horizon.

Remarkably,  such  a  no-go   mechanism  may  already  exist  in  {\em
classical}   general   relativity.   In   this   scenario,  what   are
conventionally considered to be  black hole candidates are proposed to
be  eternally  collapsing objects  (ECOs)  \cite{mitra}, with  various
initial  mass  distributions  lead  to  $m=0$  eventually  (cf.   Ref.
\cite{mitra}    and   references    therein).    Further,    in   Ref.
\cite{tanmay1},   the  inherently  quantum   functional  Schr\"odinger
formalism applied to the quantum  collapse of and radiation from a 3+1
dimensional shell  of matter  finds that the  event horizon  may never
form  in finite time.   In addition,  the radiation  as seen  from the
outside   observer's   perspective  turns   out   to  be   non-thermal
\cite{tanmay2}, which allows for information to be emitted out.

There is some  supporting observational evidence in terms  of a quasar
containing intrinsic magnetic  moment \cite{schild}, indicative of the
absence of an event horizon.

Conventional belief in the existence of black holes rests on the exact
Oppenheimer-Snyder  solution  to  the general  relativistic  spherical
collapse  equations  \cite{opsny}, in  which  the  collapsing body  is
modelled as ``pressureless dust", and is shown to form a black hole in
finite proper time that goes as $m^{-1/2}$.  The point of departure to
ECOs is to  note that this solution, in which  the collapsing fluid is
implicitly  of zero  internal  energy  and zero  heat  flux, does  not
correspond to  any physical fluid,  whereas in a  realistic situation,
the  collapsing fluid  will have  finite pressure  and  finite density
gradient.   In  particular, radiation  density  and  heat flux  should
increase sharply  as the  horizon is approached,  and the  build-up of
gravitationally  trapped  radiation  pressure  is  predicted  to  keep
slowing down the collapse  as the object becomes sufficiently compact.
Accordingly,  ECO  theory  predicts  that  massive  objects  suffering
spherical  gravitational collapse  never actually  form  non-zero mass
black  holes.  In  evolving towards  a black  hole, an  ECO  burns its
entire mass  into radiation,  so that the  black hole  that eventually
forms  at   infinite  proper   time,  has  mass   $m=0$  \cite{mitra}.
Obviously,  such a  Schwarzschild black  hole that  lacks  any entropy
would resolve the BHIP information loss problem.

In situations where ECO's are shown to be unavoidable, they could be a
manifestation  of  G\"odel  incompleteness  in  the  following  sense:
acceptance  of ECOs  leads us  to the  situation that,  even  though a
Schwarzschild  black  hole  occurs   as  a  solution  to  the  general
relativity  field equations, there  are no  initial conditions  on $M$
that can collapse  into the black hole in  finite time.  The existence
of such a dynamically unattainable solution can furnish a G\"odel-like
incompleteness  corresponding  to the  theory's  consistency. This  is
analogous to  the expressibility of  an unprovable proposition  in the
formal  system $P$, assumed  to be  consistent.  ECO's  then may  be a
purely classical  effect that anticipates the quantum  effect of black
hole  evaporation,  just as  the  classical  black  hole area  theorem
\cite{bek98} anticipates the notion of black hole temperature.

To clarify the  formal character of the incompleteness,  we may regard
the  collection of  physical bodies  such as  $M$ as  a  formal system
representing  the  deductive theory  of  general  relativity (just  as
Turing machines  or physical  computers may serve  as a  formal system
representing arithmetic).   By direct physical  evolution, this formal
system  `proves'  theorems  of  general relativity.   More  precisely,
physical evolution drives a  celestial body into various states, which
can  be  interpreted  metatheoretically  as representing  theorems  in
general relativity, just as a  series of symbolic manipulations in $P$
produces  new  formulas,  which  can  be interpreted  as  theorems  in
arithmetic.   We  know  `metatheoretically',  by  direct  mathematical
insight,  that a  Schwarzschild solution  of finite  mass  $m$ exists.
However, ECO theory implies that  our formal system is constrained not
to find  this out in  finite time.  G\"odel incompleteness  would then
correspond to the situation that the semiclassical Schwarzschild black
hole  is a true  solution that  cannot be  detected by  any consistent
formalization of the semiclassical theory of gravity.

A  formalized  proof of  the  existence of  black  holes  would be  an
interpretation  of the  formation  of a  Schwarzschild  black hole  in
finite  time within  the semiclassical  theory, which  would  make the
formalization  inconsistent through  BHIP.  This is  analogous to  the
situation that, according to G\"odel's first incompleteness theorem, a
proof of  proposition (\ref{eq:God}) within $P$ would  make the formal
system inconsistent through self-reference.

If the  ECO scenario  holds true generally,  and is not  restricted to
isolated bodies that collapse with (approximately) spherical symmetry,
this would offer  further support for the view  that the semiclassical
approach is consistent  but incomplete.  The eternal-collapse scenario
potentially provides  the most conservative resolution  of BHIP, since
no new physics would be  needed.  If the BHIP localization problem can
be shown to arise in the formation of any horizon, one could `predict'
ECO's as a generic consequence of the consistency of the semiclassical
theory of gravity.

However, if it turns out that there are certain mass distributions for
which the  finite proper time  formation of non-zero mass  black holes
cannot  be avoided,  we would  be led  to conclude  that semiclassical
gravity  is  probably  inconsistent,  and  new physics,  such  as  the
possible alternatives discussed below, may have to be invoked in order
to resolve BHIP.

\subsection{Consistency through selection of an unambiguous future}

Unlike  the approach  in the  preceding subsection,  the  present one,
which implements  option (B), involves  matter passing into  the black
hole, and  thus the  BHIP localization problem  is unavoidable  in the
events  pertaining to  the initial  phase  of the  infall through  the
horizon.  However, one may restore consistency at late time events, by
having  the information localizations  in the  Alice and  Bob versions
somehow   merge  eventually.   Although  this   does  not   avert  the
inconsistency in  toto, we may be  satisfied with restricting  it to a
finite measure  and avoiding the prospect  of `eternal inconsistency',
in which the two versions diverge forever.

There seems to be little freedom to alter the Alice version during the
events of the  initial phase of infall through the  horizon of a large
black hole, since  \mbox{(semi-)classical} physics presumably holds at
those events. Relatively speaking, there  is some room to maneuver the
Bob version, depending on the theory of quantum gravity.  Accordingly,
three  broad scenarios  of  late-time resolution  of the  localization
problem are  available.  First  is: (a) that  the event  $\eta$ occurs
eventually  after the  breakdown of  the  semiclassical approximation,
with both perspectives being agreed thereafter that the information is
localized inside the black hole, which, in the Bob perspective, is now
a black hole remnant or a naked singularity.

An alternative possibility is that the event $\eta$ does not occur, as
in the string  theoretic description.  This would have  meant that the
Alice and Bob perspectives remain `eternally incompatible'. Therefore,
one option is: (b) nonlocally  transferring the information as seen in
the  Alice perspective  to  the  Hawking radiation.   As  for Bob,  he
perceives the  information of the  infalling object somehow  pass into
the  Hawking radiation  without  going through  the horizon.   Another
option is:  (c) eventually destroying the information  in both Alice's
and  Bob's  perspectives. This  would  make  the localization  problem
redundant, but at the cost of unitarity.

Simple  examples of the  above three  scenarios (a),  (b) and  (c) are
presented  in  the  following   three  headings  in  this  Subsection,
respectively.  They  had originally  been proposed in  connection with
the  BHIP  information  loss  problem, rather  than  the  localization
problem.  Here  we point out that  they can be adapted  to address the
latter  at late-time  events.  In  (b) and  (c), we  expect  that this
enforcement of  consistency will produce  G\"odel-like incompleteness,
as indeed confirmed below. \\

\paragraph{The information is localized in a naked singularity or
black hole remnant.} ~\\

In the  first example, which illustrates scenario  (a), during Hawking
radiation the black hole retains  all the initial matter together with
negative energy  quanta entangled  with the Hawking  radiation without
mutual annihilation \cite{hrv05}.  As  the black hole's aggregate mass
drops  to zero  in the  semiclassical limit,  its  Hawking temperature
($\propto 1/m$) rises to infinity.   It is further assumed that such a
zero-mass,  information-bearing object  is somehow  without detectable
impact on low-energy experiments \cite{pre93}. Crucially, it is argued
that the horizon  does not vanish but recedes to  $r=0$.  To see this,
consider the Schwarzschild metric
\begin{equation}
\label{eq:schs}
ds^2 = \left(1-\frac{2m}{r}\right)dt^2 - 
\left(1-\frac{2m}{r}\right)^{-1}dr^2 - r^2d\Omega^2,
\end{equation}
where we  have used natural units  in which $G=c=1$.   At first sight,
the metric when the mass $m$ drops to zero seems to correspond to flat
spacetime,  endowed  with Minkowski  metric  $ds^2  =  dt^2 -  dr^2  -
r^2d\Omega^2$.  However, a more careful calculation shows
\begin{equation}
\lim_{r\mapsto 0^+} g_{00}(r=2m)=0;\hspace{0.5cm}
\lim_{r\mapsto 0^+} g_{rr}(r=2m)=\infty.
\end{equation}
This  corresponds  to  a  kind  of an  `informationally  dense'  naked
singularity, with  horizon at $r=0$.  Information about  $M$ exists in
the full entangled state  encompassing the singularity and the Hawking
radiation. This provides a resolution to the information loss problem.

In  regard   to  the  localization   problem,  there  is   an  initial
incompatibility  between  the   Alice  and  Bob  perspectives  because
$\epsilon$  occurs   at  close  to  $r=2m$,   whereas  $\eta$  happens
presumably at about  $r=0$, when the evaporating black  hole's size is
of  the  order of  Planck  length.  Thereafter  an unambiguous  future
localization,  and hence  restoration  of consistency,  is assumed  to
occur, with  unequivocal agreement between both  perspectives that the
information  is localized at  the quantum  black hole  singularity.  A
detailed  quantum  gravity  treatment  should presumably  replace  the
singularity with a black hole remnant. \\

\paragraph{The information becomes localized in the Hawking radiation.} ~\\

The  second  example,  exemplifying  scenario  (b),  is  based  on  an
interesting recent  proposal to reconcile  the unitarity of  the black
hole S-matrix with  Hawking's semiclassical arguments \cite{hormal03}.
It aims  to resolve  the BHIP information  loss problem by  imposing a
final-state boundary condition at the spacelike singularity inside the
black hole,  which causes  the information inside  it to  be `ejected'
into the Hawking radiation.  Here we will note that it also reconciles
the  perspectives of  Alice and  Bob, because  the information  is now
unambiguously localized-- outside the black hole.

The final state boundary condition imposed at the singularity requires
the quantum  state of $H_{M} \otimes  H_{\rm in}$ to  be the maximally
entangled state \cite{gott03}
\begin{equation}
_{M+{\rm in}}\langle\Phi|(S \otimes I)
\label{eq:finsta}
\end{equation}
where $_{M+{\rm  in}}\langle\Phi| = N^{-1/2}\sum_{j=1}^{N} {_M}\langle
j|  _{\rm  in}\langle  j|$,  $S$  is  a  unitary  transformation,  and
$\{|j\rangle_M\}$ is  an orthonormal  basis for $H_M$.   The effective
transformation from $H_M$ to $H_{\rm out}$ is seen to be \cite{gott03}
\begin{equation}
T \equiv~ _{M + {\rm in}}\langle\Phi|(S \otimes I)|\Phi\rangle_{\rm in
+ out} = \frac{1}{N}S,
\label{eq:T}
\end{equation}
the  effectively  unitary,  black  hole S-matrix.   The  $1/N$  factor
accounts  for   post-selection  and  indicates   that  a  conventional
measurement would  have resulted in the  final state (\ref{eq:finsta})
with  probability $1/N$.   This process  may be  viewed as  a  sort of
quantum  teleportation  \cite{qtele}  that  consumes the  Unruh  state
entanglement  in  order to  nonlocally  transmit  $M$'s  state to  the
outgoing  Hawking  radiation, but  without  the concomitant  classical
communication.  This  enables the  two versions to  agree on  the late
time localization of the information.

The application  of this  picture to a  semiclassical theory  of black
hole evaporation in which $\eta$ does not happen, could be as follows.
For  concreteness,  we consider  a  string-theory-like description  of
black hole  evaporation.  In Bob's perspective,  the information about
$M$ does not pass through the horizon, but is somehow transferred into
the Hawking  radiation after being  disrupted at the  heated membrane.
In Alice's  perspective, the  information does initially  pass through
the  horizon, after  which the  projection into  the black  hole final
state `teleports'  the information into the  Hawking radiation.  Thus,
an  unambiguous  future  eventually  appears,  providing  a  late-time
resolution   to  the   localization  problem.    This   imposition  of
consistency produces a G\"odel-like incompleteness.

The event  of projection of the  state in $H_M \otimes  H_{\rm in}$ to
$_{M  + {\rm in}}\langle\Phi|(S  \otimes I)$  occurs after  $M$ enters
into the  horizon, as seen from Alice's  perspective. However, because
in Bob's perspective $M$ does not enter the black hole, the scattering
process  characterized by  $T$  would  have to  be  attributed in  his
perspective to some fundamentally indeterminate quantum gravity effect
at the  heated membrane, that  destroys $M$, recreating it  in $H_{\rm
out}$.  We identify G\"odel  incompleteness with this inability of the
theory to  provide a detailed external-based description  of the black
hole S-matrix.\\

\paragraph{The information is destroyed.}  ~\\

The third example, a concrete instance  of scenario (c), is based on a
careful  critique \cite{gott03} of  the above  black hole  final state
proposal, where  it is pointed  out that departures from  unitarity of
$T$ can arise due to  interactions $U$ between the collapsing body and
the infalling part of the Hawking radiation. Because of the black hole
final  state,  the  resulting  loss  of  information  outside  is  not
compensated for  by any information  available inside the  black hole.
This  automatically  reconciles the  perspectives  of  Alice and  Bob,
because  now that  the information  is  destroyed, 
obviously   both  can   assert   the   loss  of   information
unparadoxically.   This provides  a late-time  resolution to  the BHIP
localization problem but a negative resolution to the information loss
problem.

According to  this proposal, the effective  modified transformation on
the infalling body is, in place of Eq. (\ref{eq:T}),
\begin{equation}
T ~\equiv~ _{M  + {\rm in}}\langle\Phi|(S \otimes I)U|\Phi\rangle_{\rm
in + out} ~\equiv~  _{M + {\rm in}}\langle\Phi|V|\Phi\rangle_{\rm in +
out}.
\label{eq:TU}
\end{equation}
If and only if $_{M + {\rm in}}\langle\Phi|V$ is a maximally entangled
state is $T$ unitary (after  renormalization).  If $V$ is chosen to be
a  maximally (dis)entangling interaction,  such as  the controlled-sum
gate   \mbox{$V|j,k\rangle  =   |j,(j+k)\mod  N\rangle$},   then  from
Eq. (\ref{eq:TU}), one has
\begin{equation}
T = \sum_n \frac{1}{N}|0\rangle\langle n|,
\label{eq:Tnon}
\end{equation}
i.e., the  state of the  outgoing radiation is  $|0\rangle_{\rm out}$,
irrespective of the incoming state $|m\rangle$ of the collapsing body.
Interestingly, since the final state  of the radiation is a fixed pure
state  $|0\rangle_{\rm  out}$, predictability  is  not  lost.  Such  a
nonunitary  black hole  evaporation would  serve as  a  novel `quantum
deletion' mechanism \cite{srienv}.

As before,  to apply this picture  to a semiclassical  theory of black
hole  evaporation in  which $\eta$  does not  happen, we  consider for
concreteness   a   string-theory-like   description  of   black   hole
evaporation.  In Bob's perspective, the information about $M$ does not
pass through  the horizon, but  is somehow disrupted and  destroyed at
the heated membrane.   As a result, the Hawking  radiation is a truely
thermal  mixture.   In  Alice's  perspective, the  information  passes
through  the  horizon,  and  remains  inside  until  the  final  state
projection.   The   interaction  $U$  impairs  the   fidelity  of  the
`teleportation' of the inside  information into the Hawking radiation,
thereby  destroying  the  information.  Thus,  an  unambiguous  future
eventually   appears,  providing   a  late-time   resolution   to  the
localization problem.

For  a reason  similar  to that  in  the preceding  scenario (b),  the
imposition   of   consistency   brings   G\"odel-like   incompleteness
corresponding to the fact that  in Bob's perspective the origin of the
process  represented by the  operator $T$  in Eq.   (\ref{eq:Tnon}) is
indeterminate.  This is because the actions of $U$ and $V$, as well as
the  final  state  projection  happen  behind the  horizon,  a  region
inaccessible to the infalling object in his perspective.\\

\section{Towards resolving BHIP via avoidance of self-reference
\label{sec:avsf}}
Finally,  under option  (C), we  first consider  the  possibility that
departure   from  standard   quantum   theory  can   help  avoid   the
self-reference that  leads to the BHIP localization  problem.  A naive
way to do  so would be to somehow `turn  off' Hawking radiation.  This
would prevent the evaporation of  the black hole, and thus ensure that
an {\em evaporating} black hole does not exist in the theory, with the
event  $\epsilon$ lying  eternally to  Bob's future,  as  in classical
general  relativity.   One  problem  with  this approach  is  that  it
requires  the  new  physics  to   apply  at  the  horizon,  where  the
semiclassical description  of spacetime  is expected to  be reasonably
valid.   This may  not  be insurmountable,  since  the suppression  of
Hawking radiation would hardly be noticeable.  Another problem is that
of  giving  a covariant  specification  of  a  condition that  forbids
pair-production near the horizon, given that there is no (necessarily)
strong  curvature or other  coordinate-invariant manifestation  of the
event horizon.  A breakdown in covariance might be one price to pay.

A  further  possibility under  option  (C)  is  that standard  general
relativity is inaccurate in the classical domain.  We will consider an
extreme realization of this  option. Since the relativity of spacetime
is essential to the localization problem of BHIP, the paradox vanishes
if space and time are  not relativistic, but are absolute, somewhat in
the  sense  of  the   philosopher  Immanuel  Kant  \cite{kant}.   Most
physicists  would  probably consider  this  approach unwarranted,  but
given the  seriousness of BHIP, we  think it worth at  least a passing
mention.

In his  theory of transcendental  idealism, Kant maintained  that time
and space are  pure intuitions and {\em a  priori} forms of intuition.
Considered  from the  empirical  perspective, they  form the  absolute
context  to objects in  experience, i.e.,  phenomenal objects  open to
scientific  study.  In this  respect, his  view of  space and  time is
Newtonian.  However,  the absoluteness is  epistemological rather than
ontological  in the  sense  that space  and  time are  not objects  of
perception, and  do not exist  for objects in  themselves.  Considered
from  the transcendental  perspective, space  and time  are  pure, and
exist  subjectively as  conditions  of knowledge,  i.e., as  cognitive
structuring  imposed on  perception by  the mind  \cite{sripen}.  Kant
also  maintained  that the  axioms  of  Euclidean  geometry were  {\em
synthetic} and  known {\em a  priori}. The former  qualification means
that they  are not true in  any logically necessary way,  and could be
denied   without  contradiction,   as  in   non-Euclidean  geometries.
Nevertheless the latter qualification  indicates that knowledge of the
axioms of geometry precedes  our experience of objects, depending only
on our  pure intuition (imaginative visualization) of  space and time.
Strictly speaking, this description applies to perceptual space rather
than  physical  space, but  Kant  may not  have  intended  them to  be
different.

It turns out that the proposition of absolute space and time is not as
difficult to implement as may at first seem.  For example, it is known
that the special theory of  relativity can be apparently reproduced in
Newtonian spacetime assuming Lorentz length contraction of metric rods
and slowing down of clocks  moving with respect to a putative absolute
rest frame.  In a model of gravitation in absolute space and time, all
events  in spacetime,  not merely  causally connected  ones,  would be
assumed to possess an absolute chronological ordering, with each event
being assigned a unique spacetime point.  Thus time paradoxes like the
BHIP localization  problem are automatically  forbidden.  For example,
in  Ref.   \cite{schm02}, a  detailed  model  of  this kind  has  been
proposed, in which  black holes are replaced by  stable frozen objects
of  the  type  discussed  in Section  \ref{sec:ever}.   Not  producing
Hawking radiation, they do not evaporate, which eliminates BHIP.

\section{Conclusions \label{sec:konklu}}
Various  attempts have  been made  to  resolve BHIP,  mostly aimed  at
understanding  how  information may  be  preserved  during black  hole
evaporation.   Here   we  focussed  on  the   problem  of  information
localization  in BHIP,  and  argued that  it  signals an  inconsistent
self-reference  in semiclassical  gravity.   This may  be regarded  as
evidence  of the inadequacy  of the  semiclassical treatment  of black
hole  evolution,  or  that  standard  quantum  mechanics  and  general
relativity are  incompatible.  To  restore consistency, we  require to
avert self-reference by modifying one  or both of the latter theories,
or introduce  (new) physics that imparts to  semiclassical gravity the
incompleteness that would correspond to imposing consistency.  Various
scenarios have been discussed under this rubric.

\appendix

\section{Incompleteness due to consistent self-reference
\label{sec:sritysk}}
Syntactic G\"odel incompleteness  arising from imposing consistency in
self-referential  systems can be  demonstrated using  simple classical
circuits. Self-reference can be introduced in a circuit by identifying
two  pieces of  information with  each other,  i.e., by  introducing a
loop.   Embedded in  spacetime,  such loops  correspond to  chronology
violation--   information   flow   through   closed   timelike   lines
\cite{tysk91}.  Such `self-referential  circuits' allow an information
carrier's  future self to  interact with  its past  self.  It  is well
known that this  can lead to logical paradoxes  of the following kind.
One imagines  travelling back in time and  preventing the time-machine
that permits us to travel  back in time from being built.  Consistency
then  demands   that  the  permitted  initial   conditions  should  be
compatible with self-referencing, in  this case, time travel. This may
be regarded as a kind of incompleteness.

\begin{figure}
\includegraphics[width=7.0cm]{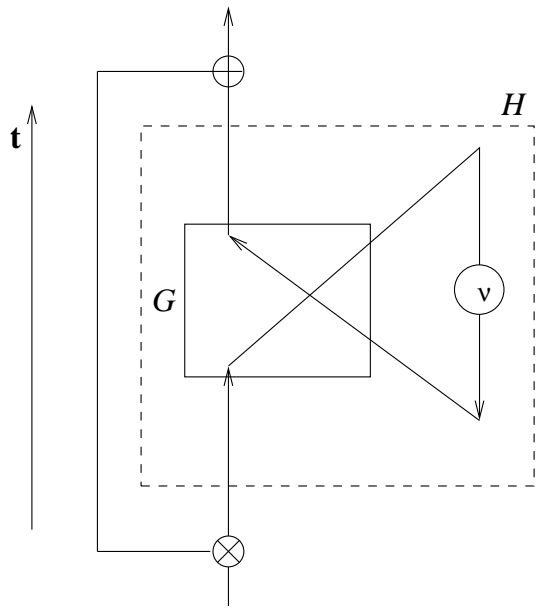}
\caption{Chronology violating subroutine  $H$, with gate $G$, embedded
in a circuit that forks the input, sending one copy through $H$, whose
output is  AND'ed with the other  copy.  The fork is  indicated by the
sign $\bigotimes$,  and AND by  $\bigoplus$.  Negative time  travel is
indicated  through  the  curve  labeled  by  the  encircled  ``$\nu$''
(Adapted from Figure 1a of Ref. \cite{tysk91}).}
\label{fig:sriganesh1}
\end{figure}

As  a simple toy  illustration, we  consider the  chronology violating
information  circuit  $H$,  outlined  by  the  dashed  box  in  Figure
\ref{fig:sriganesh1}.  The  encircled ``$\nu$'' represents  a negative
time delay whereby a bit travels back in time. The two versions of the
bit passing through the gate $G$ undergo the interaction \cite{tysk91}
\begin{equation}
G |x\rangle|y\rangle = |x\oplus y\rangle|y\rangle,
\label{eq:g}
\end{equation}
where the operation in the first  ket is XOR (addition modulo 2).  The
first and second kets refer to  the past and future selves of the bit,
respectively.  The past  self is in the state  $|x \oplus y\rangle$ on
leaving $G$, and the future self  is still in the state $|y\rangle$ in
which it entered  $G$.  That the bit does not  evolve outside the gate
imposes the consistency condition \cite{tysk91}
\begin{equation}
x \oplus y = y.
\label{eq:consi}
\end{equation}
In this case, the input value $x$ to $G$ must be zero, for no value of
$y$  satisfies the  consistency condition  (\ref{eq:consi})  if $x=1$.
This restriction due to the  demand of consistency, on what inputs may
be   processed,  may   be  regarded   as  the   system's  G\"odel-like
incompleteness.  The condition  (\ref{eq:consi}) does not uniquely fix
the  output  from  $G$,  a  problem  that may  be  amended  by  adding
supplementary   conditions   to    the   chronology   violating   bit.
Alternatively, we can  embed $H$ as a subroutine  in a larger circuit,
as  in  Fig.  \ref{fig:sriganesh1},  which  is  insensitive to  this
ambiguity.  Here one evaluates the logical  AND of a copy of the input
bit  with the  output processed  by $G$.   The demand  for consistency
retrospectively  prohibits  an  input  $x=1$  to  the  fork  from  the
unambiguous past.   However, an input  $x=0$ yields the  unique output
$y=0$  at the  unambiguous  future.   Thus the  unary  gate of  Figure
\ref{fig:sriganesh1} implements  the partial function  $\phi$ on one
bit, given  by $\phi(0)=0$ and  $\phi(1)$ being undefined.  Neither is
the full gate universal nor does it require one (NAND or NOR).

This toy model of incompleteness is meaningful only if $x=0$ and $x=1$
are  interpreted as  strictly incompatible  possibilities.  If  on the
contrary, they are interpreted as basis states of a quantum observable
(which  are only  classicaly incompatible),  and $G$  as  the unitary,
controlled-not operation $\sum_{x,y} |x\oplus y,y\rangle\langle x,y|$,
one  can avoid  the restrictions  on the  input by  allowing arbitrary
mixed-state  superpositions in  the output  $|y\rangle$ \cite{tysk91}.
The  reason is  that the  consistency condition  is equivalent  to the
requirement that an evolutionary operator have a fixed point.  Whereas
in a quantum system, there is  always such a fixed point (cf.  below),
in general, a classical system does not have such a fixed point.

The  question of  whether BHIP  may find  a solution  in the  space of
superpositions of the $\epsilon$- and $\eta$-worldlines must surely be
answered  in  the  negative,  for  the  reason  clarified  in  Section
\ref{sec:bhiloc}.  Likewise,  the outputs  $y=0$ and $y=1$  are indeed
incompatible  in  the  purely  mathematical problem  of  evaluating  a
partial function  over integers, represented by the  circuit in Figure
\ref{fig:sriganesh1}.    Thus,  our  illustration   of  G\"odel-like
self-reference  by  a  {\em  classical}  self-referential  circuit  is
appropriate.

As a circuit  version of the liar's paradox, we  consider the gate $G$
in Figure \ref{fig:sriganesh1} given by \cite{tysk91}
\begin{equation}
G |x\rangle|y\rangle = |y\oplus 1\rangle|x\rangle.
\label{eq:gp}
\end{equation}
The first and second  kets correspond, respectively, to two statements
$X$ and $Y$. The action of the  gate is to indicate the content of the
statements, with  $X$ asserting  the falsehood (NOT)  of $Y$,  and $Y$
asserting the  truth of  $X$.  This self-reference  does not  have the
earlier temporal  context.  Eq.  (\ref{eq:gp})  yields the consistency
condition \cite{tysk91}
\begin{equation}
y \oplus 1 = y,
\label{eq:pconsi}
\end{equation}
which is tantamount to requiring  the NOT gate being equivalent to the
identity operation. This cannot be  satisfied, and hence rules out all
inputs. (A quick  way to see why a consistent  quantum solution is not
ruled  out,  is  to note  that  the  unitary  version  of $G$  in  Eq.
(\ref{eq:gp}) acts  as an identity operation  between the `unambiguous
past'  and  `unambiguous  future'   selves  of  the  qubit,  with  the
chronology  violating qubit  being described  by any  density operator
$\rho  = \frac{1}{2} [\hat{\mathbb{I}}  + \zeta(|0\rangle\langle  1| +
|1\rangle\langle 0|)]$  that is  a fixed point  of the  NOT (bit-flip)
operation,  that  is, $\rho  =  \hat{X}\rho$.)


\begin{thebibliography}{100}
\bibitem{semantic}  From a  semantic  standpoint, a  formal system  is
consistent if only true propositions are provable, and complete if all
true  propositions are  provable.   The following  hold \cite{usp}:  a
semantically consistent  (complete) system is  syntatically consistent
(complete). If a formal system  is semantically consistent, then it is
semantically complete if and only if it is syntactically complete.
\bibitem{Omega}  A system  is $\omega$  consistent if  it  contains no
formula $f$ such that  it can be proved in the system  to hold for any
specific agrument, and  yet the negation of the  generalization of $f$
over  all arguments is  also provable.   Clearly, $\omega$-consistency
implies consistency.  The converse is not true.
\bibitem{add} Adding  $[R(q);q]$ or $\neg[R(q);q]$ as an  axiom to $P$
makes  Eq. (\ref{eq:God})  trivially  provable or  refutable, but  the
resulting,  augmented axiomatic system  $P^{\prime}$ will  contain its
own  G\"odel  sentence  (and,  in  the latter  case,  be  semantically
inconsistent).
\bibitem{re}  That is,  there  is  an algorithm  for  listing all  set
members,  in  the sense  that  any member  in  the  set is  eventually
detected in finite time.
\bibitem{rec} A set  is recursive if and only if both  the set and its
complement  are  r.e. Thus,  there  is  a  membership algorithm  which
determines in finite  time whether a given element  belongs to the set
or not.
\bibitem{halt}  In  computability  theory  the halting  problem  is  a
decision problem  which can be  informally stated as follows:  Given a
description of a  program $p$ and a finite  input $i$, decide whether,
given  that  input, the  program  will  eventually  halt or  will  run
forever. The halting  function $h(p,i) = 1$ if $p$  halts on $i$, else
$h(p,i)=0$.
\bibitem{2sys}  A  purported entanglement  that  can  account for  the
incompatibility    should   presumably    have    the   form    $|{\rm
Alice}\rangle|\epsilon\rangle +  |{\rm Bob}\rangle|\eta\rangle$, which
is  meaningless  in quantum  mechanics  since  the  two terms  in  the
superposition come from two  different Hilbert spaces, and makes sense
only  if  Alice  and Bob  are  two  states  of  the same  system  (or,
alter-egos of the  same person, as it were), which  is clearly not the
case here, since they represent  two distinct observers, and hence two
distinct systems.
\bibitem{go2}  G\"odel's second incompleteness  theorem shows  that if
$P$ is consistent, then its  consistency is unprovable within $P$.  In
outline, the proof is  as follows.  Metamathematically, from G\"odel's
first theorem it follows that consistency of $P$ implies unprovability
of proposition (\ref{eq:God}).  If  $w$ is the sentence that expresses
the  consistency of  $P$  in $P$,  then  the above  conclusion can  be
formalized   within  $P$   to   express  the   notion  ``$w$   implies
(\ref{eq:God})'', since (\ref{eq:God})  asserts its own unprovability.
Thus a proof  of $w$ within $P$ would yield  a proof of (\ref{eq:God})
within $P$,  contradicting G\"odel's first  incompleteness Theorem. If
$P$ is  inconsistent, then  any proposition expressible  in it  can be
proven, including its own consistency.
\bibitem{sripen} This would seem  to require that conscious perception
itself is  not a conventional objective physical  phenomenon, which is
agruably in  line with  the view  of S. Penrose  ({\em Shadows  of the
Mind}, Cambridge, 2000), who argues that it is not algorithmic.
\end{thebibliography}
\end{document}